\documentstyle[emulateapj]{article}

\slugcomment{To appear in PASP, 2006 March}

\begin{document}

\title{The Light Echo Around Supernova 2003gd in Messier 74\footnote{Based in
part on observations with the NASA/ESA {\sl Hubble Space Telescope}, obtained
at the Space Telescope Science Institute (STScI), which is operated by AURA,
Inc., under NASA contract NAS5-26555.}}

\author{Schuyler D.~Van Dyk}
\affil{Spitzer Science Center, Caltech, Mailcode 220-6, Pasadena CA  91125}
\authoremail{vandyk@ipac.caltech.edu}

\author{Weidong Li and Alexei V.~Filippenko}
\affil{Department of Astronomy, 601 Campbell Hall, University of
California, Berkeley, CA  94720-3411}
\authoremail{weidong@astro.berkeley.edu, alex@astro.berkeley.edu}

\begin{abstract}
We confirm the discovery of a light echo around the Type II-plateau Supernova
2003gd in Messier 74 (NGC 628), seen in images obtained with the High
Resolution Channel of the Advanced Camera for Surveys on-board the {\sl Hubble
Space Telescope\/} ({\sl HST}), as part of a larger Snapshot program on the
late-time emission from supernovae.  The analysis of the echo we present
suggests that it is due to the SN light pulse scattered by a sheet of dust
grains located $\sim$113 pc in front of the SN, and that these grains are not
unlike those assumed to be in the diffuse Galactic interstellar medium, both in
composition and in size distribution.  The echo is less consistent with
scattering off carbon-rich grains, and, if anything, the grains may be somewhat
more silicate-rich than the Galactic dust composition.  The echo also appears
to be more consistent with a SN distance closer to 7 Mpc than 9 Mpc.  This
further supports the conclusion we reached elsewhere that the initial mass for
the SN progenitor was relatively low ($\sim 8$--9 $M_{\odot}$).  {\it HST\/}
should be used to continue to monitor the echo in several bands, particularly
in the blue, to better constrain its origin.
\end{abstract}


\section{Introduction}

The scattering of supernova (SN) light by dust nearby to the event in the host
galaxy is likely a common occurrence.  The presence of light echoes around
supernovae (SNe) has been inferred based on infrared excesses (e.g., Dwek 1983;
Graham et al.~1983;  Graham \& Meikle 1986).  However, up until recently, only five 
SNe have had echoes unambiguously discovered around them: SN 1987A in the Large
Magellanic Cloud, SN 1991T in NGC 4527, SN 1993J in Messier 81 (M81), SN 1998bu
in Messier 96 (M96), and SN 1999ev in NGC 4274 (Maund \& Smartt 2005).  In the
cases of the Type Ia SNe 1991T (Schmidt et al.~1994) and 1998bu (Cappellaro et
al.~2001), indications of a light echo were evident in the ground-based,
late-time optical observations.  However, it required the superior angular resolution
of the {\sl Hubble Space Telescope\/} ({\it HST}) to visually confirm for both
SNe the existence of the echoes (Sparks et al.~1999; Cappellaro et al.~2001).
Although for the Type II SN 1987A the interstellar (e.g., Crotts 1988) and
circumstellar (e.g., Emmering \& Chevalier 1989; Bond et al.~1990) echoes could
be discovered from the ground, for the Type IIb SN 1993J the discovery of
echoes (Sugerman \& Crotts 2002; Liu, Bregman, \& Seitzer 2003) again is a
result of the high-resolution imaging capabilities of {\sl HST\/}.

These light echoes result as the luminous ultraviolet (UV)/optical emission
pulse from a SN is scattered by dust in dense regions of the SN environment.
The UV pulse will tend to photoionize the circumstellar matter and destroy
smaller dust grains nearest to the SN, while more distant, larger grains
survive the pulse; SNe therefore have the potential to illuminate the most
distant interstellar material and the largest structures in the environment
(Sugerman 2003).  We observe the echo as a ring, or arc, but it is actually an
ellipsoid with the SN and the observer at the foci and defined by the light
travel time from the SN (see, e.g., Fig.~1 in Patat 2005).  Light echoes
provide a means to probe both the circumstellar and interstellar structures
around SNe.  With a precise distance to the SN and the observed geometry of the
echo, we can accurately determine the three-dimensional distribution of dust in
the SN environment.

Conversely, as was elegantly shown by Panagia et al.~(1991) in the case of SN
1987A, light echoes around a SN provide a means to measure the distance to the
SN, based purely on geometrical arguments and independent of any distance
ladder.  The polarized light from the dust echo may facilitate this distance
determination (Sparks 1994, 1996).  Finally, knowing the SN spectrum, which is
what is scattered by the dust echo, we can determine the size distribution and
composition of the dust (e.g., Sugerman 2003).
 
Sugerman (2005) has recently discovered and analyzed a light echo around the Type
II-plateau (II-P) SN 2003gd in Messier 74 (M74), seen in {\sl HST\/} images we
obtained when the SN was appreciably fainter.  Here we confirm the discovery of
the echo and provide a different analysis.

SN 2003gd was discovered by Evans (2003) on 2003 June 12.82 (UT dates are used
throughout this paper) and, based on the light-curve plateau, Van Dyk, Li, \&
Filippenko (2003) estimate the explosion date at about 2003 March 17.  SN
2003gd is a somewhat unusual SN II-P and was recently discussed in detail by
Hendry et al.~(2005).  Of notable interest is that both Van Dyk et al.~(2003)
and Smartt et al.~(2004) independently determined, using a combination of
pre-SN {\sl HST\/} and ground-based images, that the progenitor was a $\sim 8\
M_{\odot}$ red supergiant (RSG), at the lower mass limit of theoretical
predictions for core-collapse SNe.  The confirmation of the progenitor star was
based on late-time {\sl HST\/} images obtained by Smartt et al.~(2004) at age
$\sim 137$~d, when the SN was slightly off the plateau, but still quite bright
in the images (see also Hendry et al.~2005).
 
\section{Observations}

In Van Dyk et al.~(2003) we presented the early-time $BVRI$ light curves for SN
2003gd, based on monitoring with the Katzman Automatic Imaging Telescope
(KAIT).  We have continued monitoring the SN with KAIT and therefore update the
ground-based light curves in Table 1.  We also list the $BR$ late-time
magnitudes from our ACS Snapshot images in Table 1.  Additionally, we have
attempted to measure the SN brightness in the ACS/HRC images obtained by Smartt
et al.~(2004) on 2003 Aug. 1; the SN is hopelessly saturated in their F814W
image, but we are able to measure F435W and F555W magnitudes for the SN through
a $0{\farcs}5$-radius aperture.  We include these magnitudes, after correction
and photometric transformation, in Table 1.

We observed SN 2003gd on 2004 December 8 with the Advanced Camera for Surveys
(ACS) High Resolution Channel (HRC) as part of our larger Cycle 13 Snapshot
program on the late-time emission from SNe (GO-10272; PI: Filippenko).  These
images were obtained when the SN was at an age $\sim 632$ d (1.73 yr), at
significantly later times than the Smartt et al.~(2004) images.  The bandpasses
and exposure times we used were F435W (840 s) and F625W (360 s).  All of the
data for this program have no proprietary period, and thus we obtained these
data from the {\sl HST\/} public archive, where standard pipeline procedures
had been employed to calibrate the images.

Unfortunately, in the F435W image a cosmic-ray hit or hot pixel sits directly
along the echo due west of the SN, such that the standard pipeline was unable
to reject this pixel from the combination of the cosmic-ray split observations.
We used the IRAF\footnote{IRAF (Image Reduction and Analysis Facility) is
distributed by the National Optical Astronomy Observatories, which are operated
by the Association of Universities for Research in Astronomy, Inc., under
cooperative agreement with the National Science Foundation.} tasks ``fixpix''
and ``epix'' to interpolate the affected pixel as well as we could.  In Figure
1 we show the corrected F435W ($\sim B$) image and the F625W ($\sim R$) image.
Although at a relatively low signal-to-noise ratio, the light echo can be
readily seen in the {\sl HST\/} images.  (The echo was not detectable in the
earlier images by Smartt et al.~or in the pre-SN {\sl HST\/} images.)  The
relatively bright object within the echo is SN 2003gd.

We measured the SN brightness in both bands, first with a $0{\farcs}5$-radius
aperture and then via point-spread function (PSF) fitting (with an equivalent
aperture also of $0{\farcs}5$ radius).  The model PSFs were constructed from
two isolated stars in the ACS/HRC images.  What is most notable is that the
aperture magnitudes for the SN are brighter than the PSF magnitudes, almost
certainly because of contamination in the aperture by the echo itself.  We
adjust the PSF magnitudes to infinite aperture, using the corrections for the
HRC in Sirianni et al.~(2005), and find $m_{\rm F435W} = 23.76 \pm 0.07$ and
$m_{\rm F625W} = 22.96 \pm 0.05$ mag.  Using the photometric transformations
also in Sirianni et al., we derive $B= 23.73 \pm 0.08$ and $R= 22.90 \pm 0.05$
mag (Table 1).  We caution that these transformations are derived from stars
with normal photospheres and not for emission, reflected or otherwise, from
sources with unusual spectra, such as SNe.  The uncertainties in $B$ and $R$
given here are strictly those in the photometric measurements and in the
transformation coefficients, and likely underestimate the actual uncertainties.

The light echo has an asymmetric structure: Only an arc of emission, most
noticeably to the northwest, not a complete ring, is seen in both the F435W and
F625W bands.  Some far weaker emission is seen to the south of the SN; the
emission to the east is not part of the echo, but instead are the stars C and D
noted by Smartt et al.~(2004).  After all the stars, including the SN, were
subtracted using the model PSFs from the images, the surface brightness of the
echo was measured in both bands.  We used the IRAF task ``minstatistics'' with
an arc-shaped pixel mask to determine the average count rate per pixel in the
echo, i.e., $0.024 \pm 0.013$ s$^{-1}$ pixel$^{-1}$ over 82 pixels in F435W and
$0.032 \pm 0.021$ s$^{-1}$ pixel$^{-1}$ over 78 pixels in F625W.  After
subtracting the average sky pixel count rate, and with the zero points from
Sirianni et al.~(2005) and a HRC plate scale of $0{\farcs}027$ pixel$^{-1}$,
these translate to average surface brightnesses of $< \mu_{\rm F435W} > = 21.5
\pm 0.5$ and $< \mu _{\rm F625W} > = 21.1 \pm 0.6$ mag arcsec$^{-2}$.
Integrating over the echo in each band we derive 
$m_{\rm F435W} = 24.5 \pm 0.5$ and $m_{\rm F625W} = 24.2 \pm 0.6$ mag,
with negligible change in the transformation (again following Sirianni et al.) 
to $m_B = 24.5 \pm 0.5$ 
and $m_R = 24.2 \pm 0.6$  mag, given the echo's color, i.e., $B-R=0.3 \pm 0.8$ mag.
Assuming Vega as photometric zero point, the echo has fluxes $1.1 \pm 0.7
\times 10^{-18}$ and $4.9 \pm 2.8 \times 10^{-19}$ erg cm$^{-2}$ s$^{-1}$
\AA$^{-1}$ at $B$ and $R$, respectively (Table 2).  Note that Sugerman (2005)
finds somewhat different values for both the surface brightness ($\mu_{\rm
F435W}=20.8 \pm 0.2$ and $\mu _{\rm F625W}=21.4 \pm 0.3$ mag arcsec$^{-2}$) and
flux ($m_B=24.2 \pm 0.1$ and $m_R=23.9 \pm 0.1$ mag), although these values
agree with ours to within the uncertainties.  The larger uncertainties we estimated for 
the fluxes, relative to those estimated by Sugerman (2005), arise from the standard
deviation in the count rate statistics within the pixel mask.  Given the low signal-to-noise 
ratio of the echo in the images in both bands, we consider our uncertainties to be quite 
conservative.

\section{Analysis}

Here we provide an analysis of the echo and its origin.  We note that this
analysis differs from that presented by Sugerman (2005).  We have determined
that SN 2003gd is at the exact center of the light echo, with uncertainty 
$< 0.2$ pixel ($< 0{\farcs}005$), through comparison of our Snapshot images to the
ACS F435W images obtained by Smartt et al.~(2004), when the SN was
significantly brighter.  The SN itself therefore must be the source of the
echo, which we observe at age $t$ after explosion and age $\tau$ after optical
maximum.  The observed echo is the product of the input SN pulse and scattering
by dust in the environment.

Following Liu et al.~(2003) and Schaefer (1987), we can approximate the
ellipsoid near the SN as a paraboloid.  The perpendicular linear distance of
the line-of-sight to the SN from the line-of-sight to the echo, the so-called
``impact parameter,'' is $b=D \theta$, where $D$ is the SN's distance from
Earth and $\theta$ is the angular distance of the two lines-of-sight.  We
measure a radius for the echo of $11.5 \pm 1.0$ pixel which then corresponds to
$\theta=0{\farcs}31 \pm 0{\farcs}03$.  For the SN distance we assume $d=7.2$
Mpc (Van Dyk et al.~2003), $b=10.8 \pm 1.1$ pc.  We note that Smartt et
al.~(2004) assume a SN distance of 9.1 Mpc and Hendry et al.~(2005) have
estimated a distance of 9.3 Mpc.  For the latter distance, $b=14.0 \pm 1.3$ pc
(hereafter, we will also provide in parentheses estimates of the various
parameters assuming a distance of 9.3 Mpc).  The age $t$, derived from the
assumed explosion date, 2003 March 17 (Van Dyk et al.~2003), is 631~d, or
1.73~yr.  However, the echo really appears due to the SN pulse, primarily in
the UV and the blue; we will assume that the SN 2003gd $B$ light curve is
similar to that of SN 1999em (see below), and Leonard et al.~(2002) determine
that the $B$ maximum for the latter SN occurred about 8 days after explosion.
Therefore, $\tau=623$ d, or 1.71 yr.  The distance from the SN to the echo,
$r=l+c\tau$, can be derived from $r^2 = b^2 + l^2$.  For $c\tau=0.52$ pc, we
find $l= 112.0 \pm 24.0$ pc and $r=112.5 \pm 23.5$ pc ($l=188.0 \pm 37.0$ and
$r=188.5 \pm 37.0$ pc for 9.3 Mpc).

Given this distance and the echo's overall asymmetric structure, the echo is
most likely from interstellar, not circumstellar, dust: For a duration of the
RSG phase $\sim 10^4$ yr and wind speed $\sim 10$ km s$^{-1}$, the
circumstellar matter would only be $\sim 0.1$ pc ($0{\farcs}09$) in radius.
Additionally, much of this dust is likely destroyed by the UV SN pulse
(Sugerman 2003).  The observed thickness of the echo is $\sim 2$ pixels, but
the stellar image width (FWHM) is larger than this, so that the echo must be
barely resolved, if at all.  Therefore, any estimate of the dust sheet
thickness along the line-of-sight (Liu et al.~2003) must be an upper limit: For
the echo, $\Delta{\theta}$ is then $\gtrsim 0{\farcs}05$, or $\Delta b \gtrsim
1.7$ pc ($\gtrsim 2.2$ pc).  The dust sheet thickness is then $\Delta l =
(b/ct) \Delta b \gtrsim 35$ pc ($\gtrsim 59$ pc).

As is generally done, we assume that the echo arises from single scattering in
a thin sheet of dust between us and the SN, and that the sheet thickness is
much smaller than the distance between the SN and the sheet. Following the
formalism of Chevalier (1986), Cappellaro et al.~(2001), and Patat (2005), the
flux $F$ at time $t$ from the echo at a given wavelength or bandpass is

\begin{equation}
F_{\rm echo} (t) = \int_0^t F_{\rm SN} (t-t^{\prime}) f(t^{\prime}) dt^{\prime},
\end{equation}

\noindent where $F_{\rm SN} (t-t^{\prime})$ is the flux of the SN at time
$t-t^{\prime}$, and $f(t)$ (in units of s$^{-1}$) determines the fraction of
light scattered by the echo toward the observer and depends on the echo
geometry and the nature of the dust.  The total SN light is effectively treated
as a short pulse over which the SN flux is constant.

The term $f(t)$ is assumed to have the form

\begin{equation}
f(t)= {{c N_H} \over r} \int Q_{\rm sca}(a) \sigma_g(a) \Phi({\alpha}, a) \phi(a) da,
\end{equation}

\noindent 
where $N_H$ is the H number density, $Q_{\rm sca}(a)$ is the scattering
coefficient for a given grain radius $a$, $\sigma_g(a)= \pi a^2$ is the dust
grain cross section for scattering, and $\Phi(\alpha)$ is the phase function
(Henyey \& Greenstein 1941)

\begin{equation}
\Phi({\alpha}, a) = {{1 - g(a)^2} \over {4 \pi [1 + g(a)^2 - 2g(a) \cos(\alpha)]^{3/2}}},
\end{equation}

\noindent 
where $\alpha$ is the scattering angle defined by
$\cos(\alpha) = [(b/c\tau)^2 - 1] / [(b/c\tau)^2 + 1]$ (e.g., Schaefer 1987).
The function $\Phi(\alpha)$ is applicable for the bandpasses being considered here
(see Draine 2003).  The term $g(a)$ measures the degree of forward scattering
for a dust grain of radius $a$.  From the geometric parameters for this echo, 
the scattering angle is then $\alpha \approx 5{\fdg}5$ ($4{\fdg}3$).
The term $\phi(a)$ is the grain size distribution for grain radius $a$.  Following
Sugerman (2003), we consider the dust grain distributions for (spherical)
silicate and carbonaceous grains from
Weingartner \& Draine (2001), and the $Q_{\rm sca}(a)$ and
$g(a)$ for ``smoothed UV astronomical silicate'' grains (Draine \& Lee 1984;
Laor \& Draine 1993; Weingartner \& Draine 2001) and for carbonaceous
graphite (Draine \& Lee 1984; Laor \& Draine 1993).

To derive the SN fluence we must integrate the light curves over time in each
band.  Unfortunately, SN 2003gd was caught late in its evolution, $\sim 90$~d
after explosion.  In Van Dyk et al.~(2003) we showed from the initial
ground-based $BVRI$ light curves that the agreement with the light curves in
the same bands for the Type II-plateau SN 1999em (Hamuy et al.~2001; Leonard et
al.~2002, 2003) is quite good on the plateau.  It is after the plateau that SN
2003gd and SN 1999em fail to agree well.  This is due to SN 2003gd being among
the peculiar, low luminosity, low $^{56}$Ni yield, SNe II-P, including also SN
1997D (Turatto et al.~1998; Benetti et al.~2001) and SN 1999br (Zampieri et
al.~2003; Pastorello et al.~2004).  However, these latter SNe all appear to
agree relatively well with SN 1999em near maximum and early on the plateau as
well.
  
In Figure 2 we show a more complete set of light curves in the $B$ and $R$
bands for SN 2003gd, which includes the updated ground-based datapoints and the
addition of the {\sl HST\/} photometry.  The SN 2003gd light curves have been
adjusted in time to match the SN 1999em light curves; no adjustment in
magnitude was necessary.  The relatively good match on the plateau phase for
both SNe suggests that we can employ the early-time SN 1999em data to
extrapolate the SN 2003gd light curves back to the date of explosion.
Performing the integration, assuming Vega as the flux zero point, we find $7.26
\times 10^{-8}$ and $8.85 \times 10^{-8}$ erg cm$^{-2}$ \AA$^{-1}$ in $B$ and
$R$, respectively (note that these differ with the fluences
reported by Sugerman 2005, i.e., $8.0 \times 10^{-8}$ and
$7.0 \times 10^{-8}$ erg cm$^{-2}$ \AA$^{-1}$ in $B$ and $R$, respectively).  
If the SN 2003gd light curves evolved in a similar
manner to SN 1999em at early times, then these fluences (and the color curves
for SN 1999em from Leonard et al.~2002) emphasize how red the SN likely became
soon after maximum light.  That is, the SN pulse resulting in the echo was
relatively red in color.

The duration of the SN pulse in each bandpass can be obtained by assuming
$F_{\rm SN} {\Delta}t_{\rm SN} = \int_0^{\infty} F_{\rm SN}(t) dt$ (Cappellaro
et al.~2001; Patat 2005).  Here we take $F_{\rm SN}$ to be the SN maximum flux,
which we assume to be the $B$ and $R$ maximum fluxes for SN 1999em (Leonard et
al.~2002), i.e., $m=13.79$ and 13.63 mag in $B$ and $R$, respectively, and
${\Delta}t_{\rm SN}$ to be the pulse duration (this is actually what is termed
the ``effective width'' of the pulse).  We then find ${\Delta}t_{\rm SN}$ to be
$\sim 45$ d in $B$ and $\sim 138$ d (about three times longer) in $R$.

We can estimate the H column density from the extinction toward the echo.  We
assume that, since the echo lies quite close to the line-of-sight to SN 2003gd,
the echo suffers the same amount of extinction as does the SN.  For the SN we
estimate a total reddening $E(B-V)=0.13 \pm 0.03$ mag (Van Dyk et al.~2003;
Smartt et al.~2004 derive a consistent estimate of the SN reddening, but with a
larger uncertainty, $E(B-V)=0.11 \pm 0.16$ mag).  Next, we must subtract the
Galactic reddening contribution, $E(B-V)=0.07$ mag (Schlegel, Finkbeiner, \&
Davis 1998).  Assuming the ratio of total-to-selective extinction $R_V = 3.1$
(e.g., Cardelli, Clayton, \& Mathis 1989), we find $A_V = 0.19$ mag internal to
the host galaxy.  Bohlin, Savage, \& Drake (1978) found a fairly constant
empirical relation over the diffuse interstellar medium in the Galaxy, $N_H=5.8
\times 10^{21} E(B-V)$, which provides a normalization for the extinction curve
$A_V/N_H = 5.3 \times 10^{-22}$ cm$^2$ (Weingartner \& Draine 2001).  From this
relation, and including our estimated uncertainty in the reddening, we derive
$N_H=3.5 (\pm 1.7) \times 10^{20}$ cm$^2$ in the dust sheet.

In Table 2 we present the fluxes $F_{\rm echo}$ in each band for several echo
models, for which we have varied the composition of the dust grains.  We have
calculated the set of models for both our distance assumption and the Hendry et
al.~(2005) distance estimate.  The first model in the set is the diffuse
Galactic dust model from Weingartner \& Draine (2001) and Draine (2003); it
assumes solar abundances with $R_V=3.1$ and the total C abundance per H nucleon
$b_C=56$ ppm, with comparable contributions of carbonaceous and silicate dust
with radii in the range 5.0 \AA\ -- 2.0 $\mu$m.  We also consider models with
$R_V=3.1$ which are either pure silicates or pure carbonaceous grains.
Finally, we consider a model with comparable silicate and carbonaceous grain
composition, but assuming $R_V=4.0$.  The results are also shown graphically in
Figure 3.

The overall agreement of the models and the observations is remarkably good.
With the various assumed model inputs, we are reproducing the observed echo
reasonably well, and this further implies that the echo likely arises from the
diffuse interstellar dust near the SN.  The uncertainties in the model fluxes
(arising mostly from the uncertainties in the echo geometrical measurements and
in our reddening estimate) are rather large, but are comparable to the
measurement uncertainties in the observed fluxes.  What we notice is that the
C+Si model agrees quite well with the observations.  The value of $R_V$ (3.0 or
4.1) has little bearing on this agreement.  The pure Si-rich dust model is also
consistent with the observations; the pure C-rich dust model, less so (although
it agrees to within the uncertainties for $d=7.2$ Mpc).  In fact, for the
larger assumed SN distance ($d=9.3$ Mpc), the pure carbonaceous dust model is
no longer consistent with the observations at either band and can be ruled out.

We note that the remaining models calculated for the larger SN distance
generally tend to underestimate the flux, although taking into account the
large uncertainties in both the observed and model fluxes, it is impossible to
rule them out entirely.  However, we tentatively suggest that the observed echo
may indicate that the actual SN distance is closer to the smaller value we
assumed in Van Dyk et al.~(2003) than the larger one determined by Hendry et
al.~(2005; and the similarly larger distance assumed by Smartt et al.~2004).
This, along with the value of $R_V$, has implications for the absolute
magnitude, and therefore the initial mass, of the SN progenitor.  A higher
$R_V$ would imply that the progenitor was, at most, $\sim 0.1$ mag more
luminous than what we estimated in Van Dyk et al.  However, the larger distance
would require the star to be $\sim 0.6$ mag more luminous, which would increase
the mass estimate by $\sim 1\ M_{\odot}$ (i.e., it would imply that the initial
mass was closer to $\sim 10\ M_{\odot}$).  The relative agreement between the
observed echo and the echo models based on the shorter distance reassures us
that our low progenitor mass estimate ($\sim 8$--9 $M_{\odot}$), although
uncomfortably near the theoretical limit for core collapse (Woosley \& Weaver
1986), is realistic.

\section{Conclusions}

We have confirmed the presence of a scattered light echo around the nearby Type
II-plateau SN 2003gd in M74.  This discovery could only have been made in
images produced with the superior angular resolution of the {\sl HST\/} ACS/HRC
at sufficiently late times for the SN.  We conclude that the echo arises from
dust in the interstellar SN environment, and our modeling (within the large
uncertainties in the observations, which further propagate into the models)
suggests that this dust, both in composition and in grain size distribution, is
not unlike dust in the diffuse Galactic interstellar medium, although it is
also possible the dust could be more silicate-rich than carbon-rich.  In fact,
our echo models tend to disfavor dust in the SN environment which is more
abundant in carbonaceous grains than silicates.  (We note that Sugerman 2005
found that the echo may arise from small carbon-rich grains.)

The models are not particularly sensitive to the value of $R_V$ (but we did not
compute models with $R_V > 4$).  However, models based on the shorter distance
to the SN that we assumed in Van Dyk et al.~(2003; 7.2 Mpc) appear to be
somewhat more consistent with the observed echo than those for the longer
distance assumed by Smartt et al.~(2004; 9.1 Mpc) and Hendry et al.~(2005; 9.3
Mpc), though the uncertainties are large. These latter two factors slightly
increase our confidence in the relatively low estimate ($\sim 8$--9
$M_{\odot}$) for the initial mass of the SN progenitor we derived in Van Dyk et
al.  

From $N_H$ and assuming a path length $L = \Delta l \approx 35$ pc for the dust
sheet, the H number density would be $n_{\rm H} \approx 7$ cm$^{-3}$.  Combined
with the extinction to the SN, this is consistent with the expectation that
light echoes likely emerge from regions with $n_{\rm H} \approx 10$ cm$^{-3}$
and $A_V \lesssim 1$ mag (Sugerman 2003).  

This echo should be further monitored with {\sl HST}, including use of
additional bands, particularly in the UV, to far better constrain the nature of
the scattering dust and the echo geometry, and to reveal further new or
evolving structures in the echo.

\acknowledgements

The work of A.V.F.'s group at UC Berkeley is supported by NSF grant
AST-0307894, as well as by NASA grants GO-10272, AR-10297, and AR-10690 from
the Space Telescope Science Institute, which is operated by AURA, Inc., under
NASA contract NAS5-26555. A.V.F. is also grateful for a Miller Research
Professorship at U.C. Berkeley, during which part of this work was completed.
KAIT was made possible by generous donations from Sun Microsystems, Inc., the
Hewlett-Packard Company, AutoScope Corporation, Lick Observatory, the National
Science Foundation, the University of California, and the Sylvia \& Jim Katzman
Foundation.  We thank the referee for useful comments.


\begin{deluxetable}{lcccccc}
\tablenum{1}
\tablewidth{6.5truein}
\tablecolumns{6}
\tablecaption{Photometry of SN 2003gd in M74}
\tablehead{
\colhead{UT date} & \colhead{Julian Date} & \colhead{$B$} & \colhead{$V$}
& \colhead{$R$} & \colhead{$I$} \\
\colhead{} & \colhead{} & \colhead{(mag)} & \colhead{(mag)} & \colhead{(mag)} & \colhead{(mag)}}
\startdata
2003 Aug 01\tablenotemark{a} &    2452853.46 &    19.17(03) &    17.41(03) &  \nodata &    \nodata  \\
2003 Aug 25 &   2452876.96 &    19.16(08)  &   17.70(04) &    16.61(02) &   16.06(03) \\
2003 Aug 31 &   2452882.97 &    19.08(12) &    17.66(03) &    16.58(02) &   16.02(03) \\
2003 Sep 06 &   2452888.96 &    \nodata &    17.68(04) &    16.62(02)  &   16.08(02) \\
2004 Dec 8\tablenotemark{b} &    2453347.95 &    23.73(08) &    \nodata  &   22.90(05) &    \nodata \\
\enddata
\tablenotetext{}{See Van Dyk et al.~(2003) for previous photometric measurements.  Note: uncertainties in hundredths of a magnitude are indicated in parentheses.}
\tablenotetext{a}{Transformed using the prescription in Sirianni et al.~(2005) from aperture magnitudes ($0{\farcs}5$ radius aperture) measured from the ACS/HRC images obtained by program GO-9733. 
The F435W magnitude = 19.23(01), and the F555W magnitude = 17.56(01).
 The F814W image of the SN is completely saturated and therefore useless.}
 \tablenotetext{b}{Transformed using the prescription in Sirianni et al.~(2005) from PSF-fitting
 photometry measured from our Snapshot ACS/HRC images (see text).  The F435W magnitude =
 23.76(07), and the F625W magnitude = 22.96(05).}
\end{deluxetable}

\clearpage

\begin{deluxetable}{lcccccc}
\tablenum{2}
\tablewidth{6.0truein}
\tablecolumns{3}
\tablecaption{Observed and Model Fluxes for the SN 2003gd Light Echo}
\tablehead{
\colhead{} & \colhead{$F_{\rm echo}(B)$} & \colhead{$F_{\rm echo}(R)$}}
\startdata
Observed & $11 \pm 7$ & $4.9 \pm 2.8$ \\
\cutinhead{Model\tablenotemark{a}\  \ with $d=7.2$ Mpc\tablenotemark{b}}
C+Si dust, $R_V=3.1$ & $8.5 \pm 4.5$ & $3.5 \pm 1.9$ \\
Pure Si dust, $R_V=3.1$ & $14 \pm 7$ & $5.1 \pm 2.7$ \\
Pure C dust, $R_V=3.1$ & $3.3 \pm 1.8$ & $2.0 \pm 1.0$ \\
C+Si dust, $R_V=4.0$ & $8.8 \pm 4.6$ & $4.4 \pm 2.3$ \\
\cutinhead{Model\tablenotemark{a}\ \ with $d=9.3$ Mpc\tablenotemark{c}}
C+Si dust, $R_V=3.1$ & $5.3 \pm 2.8$ &  $2.2 \pm 1.1$ \\
Pure Si dust, $R_V=3.1$ & $8.5 \pm 4.5$ & $3.1 \pm 1.6$ \\
Pure C dust, $R_V=3.1$ & $2.1 \pm 1.1$ & $1.2 \pm 0.6$ \\
C+Si dust, $R_V=4.0$ & $5.4 \pm 2.9$ & $2.7 \pm 1.4$ \\
\enddata
\tablenotetext{}{Fluxes are in $10^{-19}$ erg cm$^{-2}$ s$^{-1}$ \AA$^{-1}$.}
\tablenotetext{a}{The standard model for dust in the Galactic diffuse ISM is
for grain radii 5.0 \AA\ -- 2.0 $\mu$m and $b_C=6 \times 10^{-5}$ (see
Weingartner \& Draine 2001).}
\tablenotetext{b}{SN distance from Van Dyk et al.~(2003).}
\tablenotetext{c}{SN distance from Hendry et al.~(2005).}
\end{deluxetable}


\begin{figure}
\figurenum{1}
\caption{{\sl HST\/} images obtained at late times using
the ACS/HRC of the SN II-P 2003gd in M74, showing the light echo
around the SN in the ({\it a}) F435W ($\sim B$) and ({\it b}) F625W 
($\sim R$) passbands.}
\end{figure}

\clearpage

\begin{figure}
\figurenum{2}
\plotone{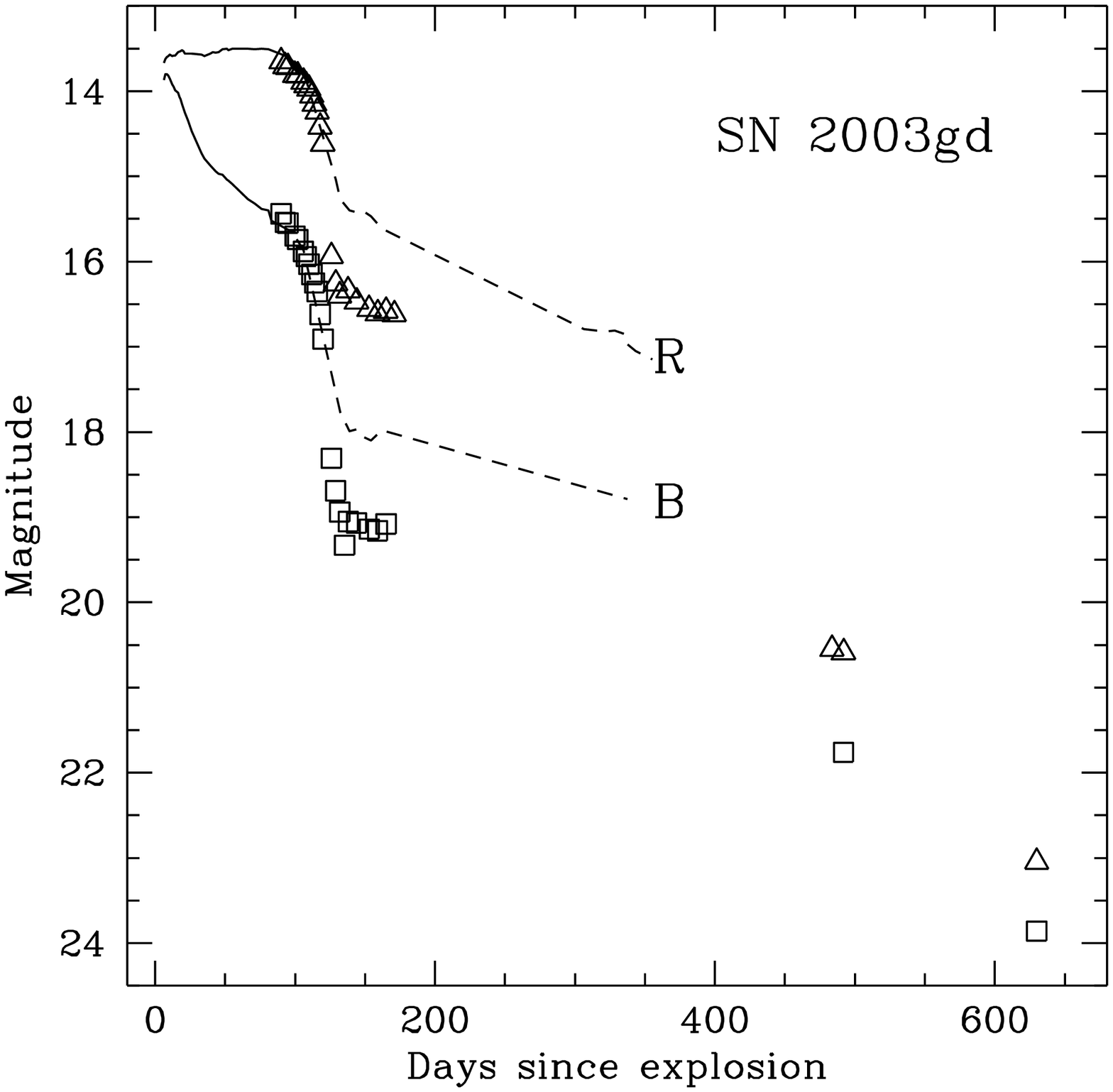}
\caption{$BR$ light curves for SN 2003gd from KAIT and {\sl HST}/ACS
observations.  For comparison, the light curves for the well-studied SN II-P
1999em (Hamuy et al.~2001; Leonard et al.~2002, 2003) are shown, adjusted to
the true distance modulus of M74.  No additional reddening correction has been
applied to the SN 1999em light curves.  We extrapolate the observed SN 2003gd
light curves to maximum light following the early-time SN 1999em curves ({\it
solid lines}); the {\it dashed lines} represent the later-time light curves for
SN 1999em.}
\end{figure}

\clearpage

\begin{figure}
\figurenum{3}
\plotone{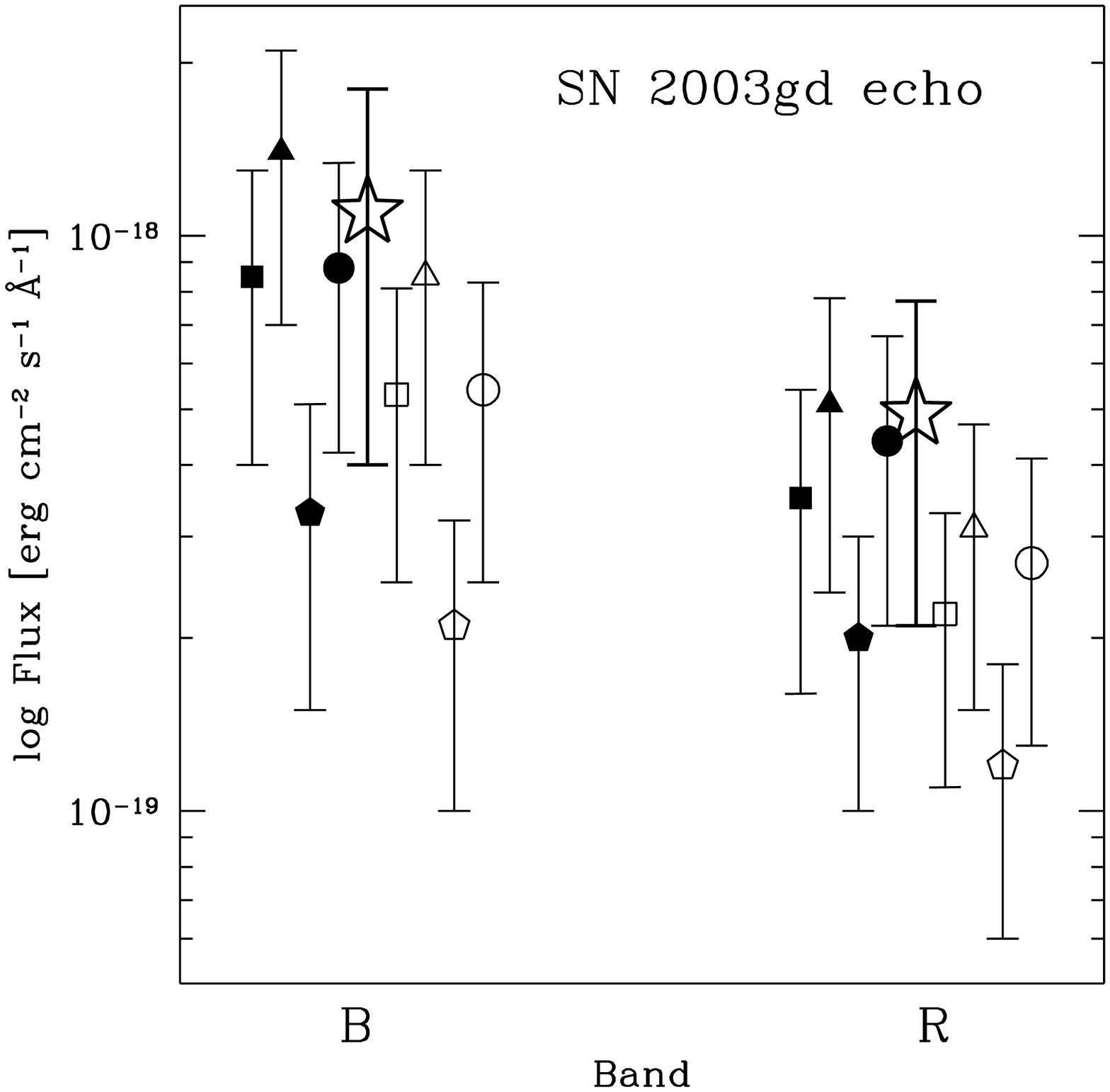}
\caption{Observed and model fluxes for the echo in the $B$ and $R$ bands (see
Table 2).  The observed fluxes are represented with {\it stars}.  The models
are represented as follows for both bands:  (1) For $d=7.2$ Mpc and $R_V=3.1$, 
C+Si, {\it filled squares}; pure Si, {\it filled triangles}; pure C, {\it filled pentagons};
and, $R_V=4.0$ and C+Si, {\it filled circles}.  (2) For $d=9.3$ Mpc and $R_V=3.1$,
C+Si, {\it open squares}; pure Si, {\it open triangles}; pure C, {\it open pentagons};
and,  $R_V=4.0$ and C+Si, {\it open circles}.  The datapoints for each band have been spread 
out along the abscissa for visual clarity.}
\end{figure}


\begin{thebibliography}{}

\bibitem[Benetti et al.~2001]{ben01} Benetti, S., et al.~2001, \mnras, 322, 361

\bibitem[Bohlin, Savage, \& Drake 1978]{boh78} Bohlin, R.~C., Savage, B.~D., \&
Drake, J.~F. 1978, \apj, 224, 132

\bibitem[Bond et al.~1990]{bon90} Bond, H.~E., Gilmozzi, R., Meakes, M.~G., \&
Panagia, N. 1990, \apj, 354, L49

\bibitem[Cappellaro et al. 2001]{cap01} Cappellaro, E., et al. 2001, \apj, 549, 
L215

\bibitem[Cardelli, Clayton, \& Mathis 1989]{car89} Cardelli, J.~A., Clayton, G.~C.,
\& Mathis, J.~S. 1989, \apj, 345, 245

\bibitem[Chevalier 1986]{che86} Chevalier, R.~A. 1986, \apj, 308, 225

\bibitem[Crotts 1988]{cro88} Crotts, A. P. S. 1988, \apj, 333, L51

\bibitem[Draine 2003]{dra03} Draine, B.~T. 2003, \apj, 598, 1017

\bibitem[Draine \& Lee 1984]{dra84} Draine, B.~T., \& Lee, H.~M. 1984, \apj, 285, 89

\bibitem[Dwek 1983]{dwe83} Dwek, E. 1983, \apj, 274, 175

\bibitem[Emmering \& Chevalier 1989]{emm89} Emmering, R.~T., \& Chevalier, R.~A. 1989, 
\apj, 338, 388

\bibitem[Evans 2003]{eva03} Evans, R. 2003, IAU Circ. 8150 

\bibitem[Graham et al.~1983]{gra83} Graham, J.~R., et al.~1983, Nature, 304, 709

\bibitem[Graham \& Meikle 1986]{gra86} Graham, J.~R., \& Meikle, W.~P.~S. 1986, 
\mnras, 221, 789

\bibitem[Hamuy et al.~2001]{ham01} Hamuy, M, et al.~2001, ApJ, 558, 615

\bibitem[Hendry et al.~2005]{hen05} Hendry, M.~A., et al.~2005, \mnras, 359, 906

\bibitem[Henyey \& Greenstein 1941]{hen41} Henyey, L.~C., \& Greenstein, J.~L. 1941,
\apj, 93, 70

\bibitem[Laor \& Draine 1993]{lao93} Laor, A., \& Draine, B.~T. 1993, \apj, 402, 441

\bibitem[Leonard et al.~2003]{leo03} Leonard, D.~C., Kanbur, S.~M., Ngeow, C.~C.,
\& Tanvir, N.~R. 2003, \apj, 594, 247

\bibitem[Leonard et al.~2002]{leo02} Leonard, D.~C., et al.~2002, \pasp, 114, 35

\bibitem[Liu, Bregman, \& Seitzer 2003]{liu03} Liu, J.-F., Bregman, J. N., \& 
Seitzer, P. 2003, \apj, 582, 919

\bibitem[Maund \& Smartt 2005]{mau05} Maund, J.~R., \& Smartt, S.~J.  2005,
\mnras, 360, 288

\bibitem[Panagia et al.~1991]{pan91} Panagia, N., et al.~1991, \apj, 380, L23

\bibitem[Pastorello et al.~2004]{pas04} Pastorello, A., et al.~2004, \mnras, 347,
74

\bibitem[Patat 2005]{pat05} Patat, F. 2005, \mnras, 357, 1161

\bibitem[Schaefer 1987]{sch97} Schaefer, B. E. 1987, ApJ, 323, L47

\bibitem[Schlegel, Finkbeiner, \& Davis 1998]{sch98} Schlegel, D.~J., Finkbeiner, 
D.~P., \& Davis, M. 1998, \apj, 500, 525

\bibitem[Schmidt et al. 1994]{sch94} Schmidt, B. P., et al. 1994, \apj, 434, L19

\bibitem[Sirianni et al.~2005]{sir05} Sirianni, M., et al. 2005, \pasp, in press

\bibitem[Smartt et al.~2004]{sma04} Smartt, S. J., et al.~2004, Science, 303, 499

\bibitem[Sparks 1994]{spa94} Sparks, W. B. 1994, \apj, 433, 19

\bibitem[Sparks 1996]{spa96} Sparks, W. B. 1996, \apj, 470, 195

\bibitem[Sparks et al.~1999]{spa99} Sparks, W. B., et al. 1999, \apj, 523, 585

\bibitem[Sugerman 2003]{sug03} Sugerman, B.~E.~K. 2003, \aj, 126, 1939

\bibitem[Sugerman 2005]{sug05} Sugerman, B.~E.~K. 2005, \apjl, in press

\bibitem[Sugerman \& Crotts 2002]{sug02} Sugerman, B.~E.~K., \& Crotts, 
A.~P.~S. 2002, \apj, 581, L97

\bibitem[Turatto et al.~1998]{tur98} Turatto, M., et al.~1998, \apj, 498, L129

\bibitem[Van Dyk, Li, \& Filippenko 2003]{van03a} Van Dyk, S. D., Li, W.,
\& Filippenko, A. V. 2003, \pasp, 115, 1289

\bibitem[Weingartner \& Draine 2001]{wei01} Weingartner, J.~C., \&
Draine, B.~T. 2001, \apj, 548, 296

\bibitem[Woosley \& Weaver 1986]{woo86} Woosley, S.~E., \& Weaver, T.~A.
1986, \araa, 24, 205

\bibitem[Zampieri et al.~2003]{zam03} Zampieri, L., et al.~2003, \mnras, 338, 711

\end{thebibliography}
\end{document}